\documentstyle[aps,preprint,12pt]{revtex}
 \begin{document}
 \draft

 \title{Effects of an electron gas on the negative trion in
semiconductor quantum wells}

 \author{Luis C.O. Dacal$^1$, Jos\'e A. Brum$^{1,2}$}

 \address{(1) DFMC-IFGW, Universidade Estadual de Campinas, C.P. 6165\\
13083-970, Campinas (SP), Brazil\\
(2) Laborat\'orio Nacional de Luz S\'incrotron - ABTLuS, C.P. 6192\\
13083-970, Campinas (SP), Brazil
}

 \maketitle

 \date{\today}

\begin{abstract}
We present here the results of calculations of the
negative trion binding energy in the presence of an electron gas.
The screening of the Coulomb interaction and the Pauli exclusion
principle are considered.  Our results show a rapid ionization
of the negative trion due to the Pauli exclusion principle while the
screening is mainly responsible for the weakening of the trion
binding energy.
\end{abstract}

\begin{center}


Corresponding author : Jos\'e Ant\^onio Brum

Address : DFMC-IFGW, Universidade Estadual de Campinas, C.P. 6165\\
13083-970, Campinas (SP), Brazil

Phone: (005519)3788-5490

Fax: (005519)3289-3137

e-mail: brum$@$ifi.unicamp.br

\end{center}

\newpage

\section{Introduction}

The optical properties of low-dimensional semiconductors are strongly
influenced by Coulomb interactions.  In the case of intrinsic quantum wells
(QWs),
 the optically created electron-hole
pair forms a bound state, the exciton, which dominates the optical
transitions \cite{chemla}.  In the presence of an electron gas, in
modulated-doped  samples, the exciton is quenched and the optical properties
are dominated by band-to-band transitions.  Coulomb correlations may
eventually manifest themselves in the optical response of the system.
This many-body effect is characterized by an enhancement of the
spectrum intensity known
as the  Fermi edge singularity \cite{skolnick}.
The situation is quite different
in the case of low-carrier densities.  In this
regime, the  electron-hole pair may still hold a bound state.  The
exciton  interacts with the electron gas (in the case of
n-doped samples) eventually binding an extra electron forming a
bound complex, the negative trion (X$^-$).  This state is the
ground state of the system while the exciton (X$^0$) is an excited
state.   In this picture, the second electron is only weakly bound since its
binding energy originates from the interaction between the electron and
a neutral exciton through its dipole.  As a consequence, the spectrum lines
associated with the X$^-$ and the X$^0$ are separated by an energy of the
order of few $meV$.  This makes the
observation of the $X^-$
a difficult task only  recently achieved in high-quality samples.
Since its first observation \cite{kheng}, the trion has been intensively
studied, both experimentally \cite{exp} and theoretically \cite{teo}.  One of the aspects that has
not been considered yet is the influence of the excess of electrons in the
trion properties.  Here we address this issue by calculating the negative trion binding
energy in the presence of an excess of electrons in semiconductor QWs.
Our results show that while the screening of the Coulomb interaction decreases
the binding energy, the phase-space filling effects originated from the Pauli
exclusion principle rapidly ionize the trion complex.

\section{Model}
\label{PC}

We consider the stability of the ground state of
two electrons and one hole complex in a  semiconductor QW in the
presence of a degenerate population of electrons  in the lowest QW
conduction subband. The trion states can be labeled by their total
spin, $S$, total angular momentum, $M$, and total wave-vector,
$K_{CM}$. In the absence of external fields, only the $M=0$
singlet ($S=0$) is a bound state.  We calculate the $K_{CM}=0$
bound state variationally by expanding the trion wave-function in
a symmetrized two-relative particles plane-wave basis
for the in-plane coordinates.  Along the growth direction
(z-direction), the component of the
wave-function is
taken as being the QW conduction and valence ground states.  The
Pauli exclusion principle is taken into account by limiting the
relative particle plane-waves to $k$ values above the Fermi
wave-vector, $k_F$.  The orbital  part of the singlet
wave-function $\Psi_{S,M,K_{CM}}$, for the ground state is written as:

\begin{eqnarray}
\Psi_{0,0,0}(\rho_1,\rho_2)=
\chi_0(z_{e1})\chi_0(z_{e2})\phi_0(z_h) \sum_{\lambda_1,\lambda_2}
\hspace{0.2cm} a_{\lambda_1 \lambda_2}
 \left\{
 \hat{\Large{\bf S}} \sum_{\vec{k}_1,\vec{k}_2 (k_1,k_2 < k_F)}
\alpha(\vec{k}_1,\vec{k}_2;\lambda_1,\lambda_2) e^{i\vec{k}_1 .
\vec{\rho}_1} e^{i\vec{k}_2 . \vec{\rho}_2} \right\}
\end{eqnarray}
where $\chi_0(z)$ and $\phi_0(z)$ are the conduction and valence
QW ground states, the $\lambda_i$'s are numerical parameters
chosen within physically meaningful values and a$_{\lambda_1
\lambda_2}$'s are the linear variational parameters. The
$\alpha(\vec{k}_1,\vec{k}_2;\lambda_1,\lambda_2)$ are chosen so
that for $k_F=0$ the wave-function is a symmetrized expansion in
non-orthogonal two single-relative-particle gaussian functions
parameterized by $\lambda_i$ is limited to the s-like relative
states \cite{kleinman}. This is enough to obtain a good
description of the trion binding energy. For a better quantitative
result it is necessary, however, to include higher relative
angular momenta.

The screening of the Coulomb interaction is considered within the
Random-Phase Approximation (RPA) \cite{brum}.  Within this
approximation, the Coulomb  interaction is in-plane Fourier transformed and
the q-dependent  Coulomb interaction is screened by the RPA static-dielectric
constant, $\epsilon_{RPA}(q)=1+\frac{2}{q a_0} F(q)g(q)
$,
where $a_0$ is the three-dimensional Bohr radius, $F(q)$ is the usual
QW form factor and
$g(q)=1-\sqrt{1-(\frac{2 k_F}{q})^2} Y(q-2k_F)
$,
where Y(x) is the step function ($Y=0$ if $x<0$ and $Y=1$ if $x>0$).  The
trion ground-state is calculated by diagonalizing the
generalized eigenvalue problem obtained by projecting the total
two-electrons and one-hole Hamiltonian into the wave-function of
Eq. 1. To obtain the conditions for the stability of the trion we
calculate the exciton plus a non-interacting electron ground state
within the same approximation.

\section{Results and Discussion}

We considered a single 100 \AA\ symmetrically n-modulated-doped GaAs-(Ga,Al)As
QW. Figure 1 shows (a) the exciton energy and (b) the
trion energy relative to the non-interacting particles total
energy as a
function of the electron  density when only the screening is
considered (dotted-line), when only the  Pauli exclusion principle
effect is considered (dashed-line) and when  both effects are
included (full line).  Here we subtracted the kinetic
energy of the relative particle with $k_F$ wave-vector for the
exciton and twice this value for the trion.
This garanties that the exciton and the trion states are not degenerate with
available scattering states and is a true bound state.  We observe
that the screening decreases the binding energy already at very
weak electron concentrations (n$_e$).  However, even for large
concentrations, it remains a sizable although weak binding energy
\cite{brum}.  On the other hand, the Pauli exclusion principle is almost
ineffective up to n$_e$ of the order of $10^9$
cm$^{-2}$.  For higher concentrations, the binding energy is
rapidly quenched \cite{kleinman} at n$_e$=$4 \times
10^{10}$ cm$^{-2}$. The combined effects show an exciton binding
energy which decreases with n$_e$ and is
quenched for a concentration of $10^{10}$ cm$^{-2}$.
  The
same behavior is observed for the trion energy, where it was
extracted twice the relative Fermi energy, following the same
arguments as in the exciton case.

The stability of the $X^-$ complex, however, must be established
by comparing its total energy with the next two-electrons one-hole
state, that is, when the one electron and the hole form the bound
excitonic state and the extra electron is at the lowest available
scattered state, that is, at the Fermi energy.  The trion binding
energy is defined as the difference between this two energies:
$E_{X^-}^b=E_{X^0}+e_F-E_{X^-}$.  In
Figure 2 we plot this value as a function n$_e$.
When we consider only the screening effects, the
trion binding energy remains roughly the same up to values of n$_e$
of the order of $5 \times 10^{10}$ cm$^{-2}$, when it increases
almost linearly with it.  The decrease in the binding energy due to
the screening is compensated by the energy of the extra electron
up to concentrations of the order of $5 \times 10^{10}$ cm$^{-2}$,
giving an almost constante trion binding energy.  Once the screening
effects is saturated, the energy increases with the Fermi energy
almost linearly.
The situation changes completly when we consider the Pauli exclusion
principle.  In this case, the trion binding energy quenches
already at values of n$_e$ of the order of $5 \times 10^{9}$ cm$^-2$.
When we considere both, the screening and the Pauli exclusion
principle, the trion ionizes at concentrations of the order of
$2 \times 10^9$ cm$^{-2}$.  In
actual samples, the nominal excess of electrons is above this
value.  However, the trion and the exciton emission lines remains stable
and no quenching of the photoluminescence is observed.
A more complex trion wave function gives higher values for the trion
binding energy and one should expect the quenching to happen at
higher concentrations.  This will not, however, change the qualitative
picture.
One of the reasons
to understand the trion stability is the effect of interface fluctuations.
It may
trap the trion, increasing the energy necessary to
ionize it\cite{dacal} or, it may trap the electrons making them less
effective in the quenching of the bound state.  More experimental and
theoretical work is necessary to elucidate this problem.

The authors are grateful to G. Bastard and B. Deveaud-Pledran for stimulating
discussions.  This work is supported by FAPESP (Brazil) and CNPq
(Brazil).


\begin{figure}
\caption{(a) Exciton binding energy and (b) negative trion binding
energy relative to the effective gap for a 100 {\AA} QW as a
function of n$_e$ when we include the effects of the screening
(dash-dotted line), the Pauli exclusion principle (dashed-line)
and when both effects are considered (full line). } \label{absexp}
\end{figure}

\begin{figure}
\caption{Trion binding energy as a function of the electron
concentration for the same cases as in Figure 1.}
\label{diagrams}
\end{figure}

\end{document}